\documentclass[11pt,a4paper]{article}
\pdfoutput=1
\usepackage{amsmath,amssymb,graphicx}
\usepackage{mathrsfs}
\usepackage{hyperref}
\usepackage{bm}
\usepackage{multirow}
\usepackage{leftidx}
\usepackage[text={17cm,24.5cm},centering]{geometry}
\usepackage{textcomp}
\usepackage{wasysym}
\numberwithin{equation}{section}

\newcommand{\be}{\begin{equation}}
\newcommand{\bea}{\begin{eqnarray}}
\newcommand{\eea}{\end{eqnarray}}
\newcommand{\ba}{\begin{array}}
\newcommand{\ea}{\end{array}}
\newcommand{\ee}{\end{equation}}
\newcommand{\ml}{\mathcal}
\newcommand{\ep}{\epsilon}
\newcommand{\de}{\delta}

\newcommand{\dg}{\dagger}
\newcommand{\ga}{\gamma}
\newcommand{\ti}{\tilde}

\newcommand{\matr}[2]{\left(\begin{array}{#1}#2\end{array}\right)}
\newcommand{\nn}{\nonumber}

\def\fR{{\mathfrak R}}

\def\({\left(}
\def\){\right)}
\def\[{\left[}
\def\]{\right]}
\begin{document}
\title{\textbf{Y-system for $\gamma$-deformed ABJM Theory}}
\author{Hui-Huang Chen$^{a, b}$\footnote{chenhh@ihep.ac.cn}~,
Peng Liu$^{a, b}$\footnote{liup51@ihep.ac.cn}~, Jun-Bao Wu$^{c, d, a, b, e}$\footnote{wujb@ihep.ac.cn}~
}
\date{}

\maketitle

\vspace{-10mm}

\begin{center}
{\it
$^{a}$Institute of High Energy Physics, and Theoretical Physics Center for Science
Facilities, Chinese Academy of Sciences, 19B Yuquan Road, Beijing 100049, P.~R.~China\\
$^{b}$University of Chinese Academy of Sciences, 19A Yuquan Road, Beijing 100049, P.~R.~China\\
$^{c}$ School of Science, University of Tianjin, 92 Weijin Road, Tianjin 300072, P.~R.~China\\
$^{d}$School of Physics and Nuclear Energy Engineering, Beihang University, 37 Xueyuan Road, Beijing 100191, P.~R.~China\\
$^{e}$Center for High Energy Physics, Peking University, 5 Yiheyuan Road, Beijing 100871, P.~R.~China}
\vspace{10mm}
\end{center}

\begin{abstract}
  We investigate the integrable aspects of the planar $\ga$-deformed ABJM theory and propose the twisted asymptotic Bethe ansatz equations. A more general method through a twisted generating functional is discussed, based on which, the asymptotic large $L$ solution of Y-system is modified in order to match the asymptotic Bethe ansatz equations. Several applications of our method in the $sl(2)$-like sector and some important examples in $\beta$-deformed ABJM are presented as well.
\end{abstract}
\thispagestyle{empty}
\newpage

\tableofcontents

\section{Introduction}
The celebrated AdS/CFT correspondence continues to be a source of exciting new results in both gauge and string theory, with the best studied example being the $\text{AdS}_5/\text{CFT}_4$ duality between four-dimensional $\ml{N}=4$  superconformal Yang-Mills theory(SYM) and Type IIB superstring theory on $\text{AdS}_5\times\text{S}^5$ \cite{Mal97}. There are also other AdS/CFT examples \cite{Bab09,Klose:2010ki} with field theories being in lower dimensions and  integrability properties discovered on both sides. One of these is the $\text{AdS}_4/\text{CFT}_3$ duality proposed in \cite{ABJM},  which relates three-dimensional $\ml{N}=6$ super Chern-Simons theory and type IIA string theory on $\text{AdS}_4\times\mathbb{CP}^3$. The integrable structure on both sides in this gauge/string duality was first studied in \cite{Minahan:2008hf}-\cite{arXiv:1009.3498}. Similarly to $\ml{N}=4$ SYM \cite{Beisert:2005fw, Beisert:2006ez}, a complete description of planar anomalous dimensions of infinitely long operators has been investigated by means of the Asymptotic Bethe
Ansatz(ABA) equations \cite{c9}. To solve the spectral problem for single-trace operators with finite length, finite size effects must be computed. The complete solution to $\text{AdS}_5/\text{CFT}_4$ spectral problem is encoded in the TBA/Y-system \cite{Gromov:2009tv}-\cite{Arutyunov:2009ur}, which has an equivalent but simpler form --- the Quantum Spectral Curve(QSC) \cite{Gromov:2013pga}. In $\text{AdS}_4/\text{CFT}_3$, the TBA/Y-system was constructed in \cite{Gromov:2009at, Bombardelli:2009xz} and recently reduced to a QSC in \cite{Cavaglia:2014exa}. One of the very impressive results is the computation of the triple wrapping corrections in ABJM theory \cite{Anselmetti:2015mda} using the QSC method. Nevertheless, the good old Y-system is still efficient for computing leading wrapping corrections.\\
  \par Integrability gives us some hopes to solve exactly highly non-trivial quantum field theories.
However, such theories are quite rare and integrable structure usually only appears in the large N limit. Notice that the integrable structures in AdS/CFT
correspondence first appear in theories with a large amount of supersymmetries.
It would be very interesting to see how far one can go by reducing the supersymmetries of the original theory while keeping integrable structure at the same time. For $\text{AdS}_5/\text{CFT}_4$, people have explored a lot through at least three approaches including the addition of flavors \cite{Chen:2004mu}-\cite{Erler:2005nr}, marginal deformations \cite{Roiban:2003dw}-\cite{Mansson:2008xv} and orbifolding \cite{c4}-\cite{deLeeuw:2012hp}. For more information on these topics see reviews \cite{Zoubos:2010kh,vanTongeren:2013gva} and references therein. Some wrapping effects in $\beta$- and $\ga$-deformed SYM theories were computed using L\"uscher method and/or Y-system in \cite{Ahn:2011xq}-\cite{Beccaria:2010kd}.  In three-dimensional case, similar aspects have been less studied. In \cite{He:2013hxd}, integrability of planar $\beta$-deformed  ABJM theories was established at two-loop order in the scalar sector.\footnote{Integrability of a $\ga$-deformation different from the ones in \cite{Imeroni:2008cr} was also studied in \cite{He:2013hxd}.} Recently, integrability of orbifold ABJM theories was studied in details in \cite{Bai:2016pxs}. \\
\par The $\ga$-deformation of $\ml{N}=4$ SYM theory are very special among classically marginal deformations in the sense that it preserves the integrable structure and can be implemented elegantly on both sides of gauge/string duality. On the gauge theory side, the deformation can be expressed through a non-commutative star product for each interaction term in the Lagrangian. On its dual string theory side,
it can be constructed by T-duality-shift-T-duality (TsT) transformations \cite{Lunin:2005jy, Frolov:2005dj}. Marginal deformations of Bagger-Lambert-Gustavsson theory
\cite{BL1}-\cite{G2}  were studied in \cite{Berman:2008be}.
Similar deformations were also studied on the gravity side in
\cite{Berman:2007tf}.  These deformations were reviewed in \cite{Berman:2007bv}. The $\ga$-deformed ABJM theories and their gravity duals were studied in \cite{Imeroni:2008cr}. Some classical string solutions in these deformed backgrounds of type IIA string theory have been studied in \cite{Schimpf:2009rk}-\cite{Ratti:2012kq}.\\
\par The integrability of $\ga$-deformed ABJM theory in \cite{Imeroni:2008cr} in the scalar sector at two-loop level can be proved in a similar way as it was done for $\beta$-deformed theory \cite{He:2013hxd}. In this paper we make the very natural assumption that the planar $\ga$-deformed ABJM theory is integrable for all the sectors and to all loop orders, and compute the twist matrix as in \cite{Beisert:2005if}. Then the proposal of asymptotic Bethe equations is straightforward to get. We investigate the duality properties which have been derived in \cite{Bai:2016pxs} carefully. We find they are all satisfied nicely as they should.
Thanks to these duality properties, a more general twisted generating functional method is proposed. To match the twisted ABA equations, a modified asymptotic large $L$ solution of Y-system is presented. Various applications have been made in the $sl(2)$-like sector. Since the $\text{AdS}_4/\text{CFT}_3$ Y-system have been proposed in \cite{Gromov:2009tv} and refined in \cite{Gromov:2009at} (see also \cite{Bombardelli:2009xz}), the investigation of its application for the non-symmetric solutions $Y_{\blacktriangleleft}\neq Y_{\blacktriangleright}$ have not been made in the weak coupling region. For the $\beta$-deformed ABJM theory, first few attempts have been made for some simple states.\\
\par The plan of the paper is the following. In section \ref{s1}, we briefly discuss the basic properties  of $\ga$-deformed ABJM theory, and the twist matrix is derived. In section \ref{s2}, we present the asymptotic Bethe ansatz for $\ga$-deformed ABJM theory. In section \ref{s3}, the twisted Y-system is proposed. Finally, in section \ref{s4} and \ref{s5} some applications are investigated.
\section{The $\ga$-deformation of ABJM theory}\label{s1}
\begin{table}\label{table1}
  \centering
  \begin{tabular}{|c|c|c|c|c|c|c|c|c|}
  \hline
  Fields&$Y^1$&$Y^2$&$Y^3$&$Y^4$&$\psi^{\dg1}$&$\psi^{\dg2}$&$\psi^{\dg3}$&$\psi^{\dg4}$\\
  \hline
  $Q^1_f$&$\frac{1}{2}$&$-\frac{1}{2}$&$0$&$0$ &$\frac{1}{2}$&$-\frac{1}{2}$&$0$&$0$ \\
  \hline
  $Q^2_f$&$0$ & $0$ & $\frac{1}{2}$&$-\frac{1}{2}$&$0$&$0$&$\frac{1}{2}$&$-\frac{1}{2}$\\
  \hline
  $Q^3_f$&$\frac{1}{2}$&$\frac{1}{2}$&$-\frac{1}{2}$&$-\frac{1}{2}$&$\frac{1}{2}$&$\frac{1}{2}$& $-\frac{1}{2}$&$-\frac{1}{2}$ \\
  \hline
  \end{tabular}
  \caption{Charges of fields under three $U(1)$ generators of $SU(4)_R$.}
\end{table}
The three-parameter deformation of ABJM theory can be performed by replacing all the ordinary products $fg$ of two fields $f$ and $g$ in the Lagrangian by the following non-commuting star product,
\be\label{star}
f*g=e^{i\pi Q_f\times Q_g}fg=e^{i\pi Q_f^{i}C_{ij}Q_g^j}fg.
\ee
We parameterize the anti-symmetric phase matrix $\textbf{C}$ by three deformation parameters $\ga_1,\ga_2,\ga_3$ as
\be
\textbf{C}=\matr{ccc}
{
0&-\ga_3&+\ga_2\\
+\ga_3&0&-\ga_1\\
-\ga_2&+\ga_1&0\\
}.
\ee
The $U(1)^3$ charges of the fundamental fields are given in the Table 1.
The gravity dual of $\ga$-deformed ABJM theory is the type IIA string theory on $\ga$-deformed background $\text{AdS}_4\times\text{CP}^3_{\ga}$, which is equivalent to the same theory on undeformed background but with twisted boundary conditions\footnote{Such relation was first obtained for type IIB string theory on $AdS_5\times S^5_\ga$ in \cite{Frolov:2005dj}.}
\be\label{TwistedBC}
\phi_i(2\pi)-\phi_i(0)=2\pi(m_i-\ep_{ijk}\ga_jJ_k),\quad m_1, m_2, m_3+\frac{m_1+m_2}{2}\in\mathbb{Z}.
\ee
Here the $\phi_i$'s are three angles on $\text{CP}^3$ and $J_i$ are the corresponding angular momentum. We parameterize the $\text{CP}^3$ by embedding it inside $\mathbb{C}^4$ as
\bea
&&Y^1=\cos\xi\cos\frac{\theta_1}{2}e^{i(+\phi_1+\phi_3)/2},\nn\\
&&Y^2=\cos\xi\sin\frac{\theta_1}{2}e^{i(-\phi_1+\phi_3)/2},\nn\\
&&Y^3=\sin\xi\cos\frac{\theta_2}{2}e^{i(+\phi_2-\phi_3)/2},\nn\\
&&Y^4=\sin\xi\sin\frac{\theta_2}{2}e^{i(-\phi_2-\phi_3)/2}.
\eea
This explains the charges in Table 1.\\
\par We must point out that the choice of the three linearly independent $U(1)$ Cartan generators of $SU(4)_R$ has some degrees of freedom. Our choice of these three charges are the same as the one in \cite{Imeroni:2008cr}, since  the three TsT transformations used there is exactly based on  $\phi_1, \phi_2, \phi_3$ directions. Our choice is not the same as the one in \cite{Caetano:2016ydc}.  However the three generators in the later paper are simply  linear combinations of the generators we used here, and the two sets of deformation parameters are also related by linear transformation. So these two choices are in fact equivalent. The three $U(1)$ charges we choose are related to the Dynkin labels $[p_1,q,p_2]$ of $SU(4)$ by $J_1\equiv J_{\phi_1}=Q^1_f=\frac{p_1}{2},J_2\equiv J_{\phi_2}=Q^2_f=\frac{p_2}{2},J_3\equiv J_{\phi_3}=Q^3_f=q+\frac{p_1+p_2}{2}$.

\subsection{The $su(4)$ sector}

The Bethe ansatz equations of $\ga$-deformed theories are the same as the ones for the undeformed theory except for adding some phases.
 In order to obtain the appropriate twist phases in the $su(4)$ sector, using the same notation as in \cite{Beisert:2005if}, we introduce the $\textbf{B}$ matrix . When permuting two scalars $Y^i$ and $Y^j$, we pick up a phase
\be
Y^i*Y^j=e^{2\pi i B_{ij}}Y^j*Y^i.
\ee
In the basis $(Y^1,Y^2,Y^3,Y^4)$, using the charges of the fundamental fields listed in Table 1, the $\textbf{B}$ matrix turns out to be
\be
\textbf{B}=\frac14\matr{cccc}
{
0&2\ga_2&\ga_1-\ga_2-\ga_3&-\ga_1-\ga_2+\ga_3\\
-2\ga_2&0&\ga_1+\ga_2+\ga_3&-\ga_1+\ga_2-\ga_3\\
-\ga_1+\ga_2+\ga_3&-\ga_1-\ga_2-\ga_3&0&2\ga_1\\
\ga_1+\ga_2-\ga_3&\ga_1-\ga_2+\ga_3&-2\ga_1&0\\
}.
\ee
$su(4)$ is a rank 3 algebra, and there are three types of excitations in this sector. We denote the three creation operators by $\ml{B}_3,\ml{B}_4,\ml{B}_{\bar4}$. Their actions are
\be
\ml{B}_3: Y^2\rightarrow Y^3, \ml{B}_4:Y^1\rightarrow Y^2,\ml{B}_{\bar4}:Y^3\rightarrow Y^4.
\ee
Since our vacuum is chosen as $\text{Tr}(Y^1Y_4^{\dg})^L$ , we now should consider the twist matrix in the basis\\
$(Y^1Y_4^{\dg}|Y_2^{\dg}Y^3,Y_1^{\dg}Y^2,Y_3^{\dg}Y^4)$. The twist matrix turns out to be
\be
\textbf{A}=\matr{c|ccc}
{
0&\ga_1-\ga_2&\ga_2-\frac12\ga_3&-\ga_1+\frac12\ga_3\\
\hline
-\ga_1+\ga_2&0&-\ga_2-\frac12\ga_3&\ga_1+\frac12\ga_3\\
-\ga_2+\frac12\ga_3&\ga_2+\frac12\ga_3&0&-\ga_3\\
\ga_1-\frac12\ga_3&-\ga_1-\frac12\ga_3&\ga_3&0\\
}.
\ee
The twist phases appearing in Bethe equations are simply given by $2\pi(\textbf{AK})_i$, where $\textbf{K}=(L|K_3,K_4,K_{\bar4})$.
This result is consistent with Appendix~\ref{appena}, where direct computations through a deformed R-matrix were performed.
For $\ga_1=\ga_2=0,\ga_3=-\beta$, this comes back to the result in \cite{He:2013hxd} for $\beta$-deformed case. \footnote{We denote the real deformation parameter in $\beta$-deformed theory as $\beta$. Notice in some previous work including \cite{He:2013hxd}, it was denoted as $\ga$ to stress that it is real.}
\subsection{The full $OSp(6|4)$ sector}
\subsubsection{The distinguished basis}
From the above example, we know the crucial point is to find out the action of the creation operators. The full $OSp(6|4)$ sector contains five types of excitations $u_1,\dots ,u_{\bar4}$, and there are five corresponding creation operators $\ml{B}_1,\dots,\ml{B}_{\bar4}$. In the distinguished grading, the actions of these creation operators on  the fundamental fields can be found as
\be
\ml{B}_1: \psi_{I+}\rightarrow\psi_{I-},\;
\ml{B}_2: Y_1^{\dg}\rightarrow\psi^{\dg 2+},\;
\ml{B}_3: Y^2\rightarrow Y^3,\;
\ml{B}_4:Y^1\rightarrow Y^2,\;
\ml{B}_{\bar4}:Y^3\rightarrow Y^4.
\ee
Then we find the twist matrix in the basis $(Y^1Y_4^{\dg}|1,Y^1\psi^{\dg 2+},Y_2^{\dg}Y^3,Y_1^{\dg}Y^2,Y_3^{\dg}Y^4)$ as
\be\label{Dtwist}
\textbf{A}'=\frac12\matr{c|ccccc}
 {
 0&0&\ga_2-\ga_1&2\ga_1-2\ga_2&2\ga_2-\ga_3&\ga_3-2\ga_1\\
 \hline
 0&0&0&0&0&0\\
 \ga_1-\ga_2&0& 0 &\ga_1-\ga_2&2\ga_2&-2\ga_1 \\
 2\ga_2-2\ga_1&0 &\ga_2-\ga_1&0& -2\ga_2-\ga_3& 2\ga_1+\ga_3 \\
 \ga_3-2\ga_2&0&-2\ga_2&2\ga_2+\ga_3&0&-2\ga_3 \\
 2\ga_1-\ga_3&0&2\ga_1&-2\ga_1-\ga_3&2\ga_3&0 \\
}.
\ee
Actually if we use a different parametrization of the $\textbf{C}$ matrix
\be
\textbf{C}=\matr{ccc}
{
 0&\de_1+2\de_2+\de_3&-\de_1 \\
 -\de_1-2\de_2-\de_3&0&\de_3 \\
 \de_1&-\de_3&0 \\
},
\ee
 we can obtain the $\textbf{A}'$ matrix more directly.
Introducing the charges
\bea
\textbf{q}_1=(0|0,1,-2,1,1)\,,\nn\\
\textbf{q}_2=(1|0,0,1,-2,0)\,,\nn\\
\textbf{q}_3=(1|0,0,1,0,-2)\,,
\eea
the twist matrix can be obtained directly as
\be\label{Trick}
\textbf{A}'=\frac12\de_1(\textbf{q}_1^T\textbf{q}_2-\textbf{q}_2^T\textbf{q}_1)
+\frac12\de_2(\textbf{q}_2^T\textbf{q}_3-\textbf{q}_3^T\textbf{q}_2)
+\frac12\de_3(\textbf{q}_3^T\textbf{q}_1-\textbf{q}_1^T\textbf{q}_3).
\ee
\subsubsection{The $\eta=-1$ grading}
In the following sections, we mainly work in the $\eta=-1$ grading. Since it is the grading that
makes easier to write the Bethe equations for the $sl(2)$ sector, it
can be called $sl(2)$-favored grading, or $sl(2)$ grading for shortness
sake, throughout the rest of the paper. In Appendix~\ref{appenb} we give the result on the the $\eta=+1$ grading (also called $su(2)$-favored or $su(2)$ grading for shortness sake). In the $sl(2)$ grading, the actions of the creation operators read
\bea
&&\ml{B}_1: Y^3\rightarrow\psi_{4-},Y^4\rightarrow\psi_{3-},
Y_1^{\dg}\rightarrow\psi^{\dg2-},Y_2^{\dg}\rightarrow\psi^{\dg1-},\nn\\
&&\ml{B}_2: Y^2\rightarrow Y^3,\psi_{3\pm}\rightarrow\psi_{2\pm},\nn\\
&&\ml{B}_3: \psi^{\dg1+}\rightarrow Y_3^{\dg},\psi_{4+}\rightarrow Y^2,
\psi^{\dg3+}\rightarrow Y_1^{\dg},\psi_{2+}\rightarrow Y^4,\nn\\
&&\ml{B}_4:Y^1\rightarrow\psi_{4+},Y^4\rightarrow\psi_{1+},
Y_2^{\dg}\rightarrow\psi^{\dg3+},Y_3^{\dg}\rightarrow\psi^{\dg2+},\nn\\
&&\ml{B}_{\bar4}:Y^2\rightarrow\psi_{3+},Y^3\rightarrow\psi_{2+},
Y_1^{\dg}\rightarrow\psi^{\dg4+},Y_4^{\dg}\rightarrow\psi^{\dg1+}.
\eea
The twist matrix is
\be
\textbf{A}=\frac12\matr{c|ccccc}
{
 0&\ga_2-\ga_1&2(\ga_1-\ga_2)&\ga_2-\ga_1&\ga_1+\ga_2-\ga_3&-\ga_1-\ga_2+\ga_3 \\
 \hline
 \ga_1-\ga_2&0&\ga_1-\ga_2&\ga_2-\ga_1&\ga_1+\ga_2&-\ga_1-\ga_2 \\
 2(\ga_2-\ga_1)&\ga_2-\ga_1&0&\ga_1-\ga_2& -\ga_1-\ga_2-\ga_3&\ga_1+\ga_2+\ga_3 \\
 \ga_1-\ga_2&\ga_1-\ga_2&\ga_2-\ga_1&0&\ga_3&-\ga_3 \\
 -\ga_1-\ga_2+\ga_3&-\ga_1-\ga_2&\ga_1+\ga_2+\ga_3&-\ga_3 & 0 & 0 \\
 \ga_1+\ga_2-\ga_3&\ga_1+\ga_2&-\ga_1-\ga_2-\ga_3&\ga_3 & 0 & 0 \\
}.
\ee
In this case the three charges are found to be
\bea
\textbf{q}_1&=&(0|1,-2,1,0,0),\nn\\
\textbf{q}_2&=&(1|0,1,-1,-1,1),\nn\\
\textbf{q}_3&=&(1|0,1,-1,1,-1),
\eea
and the twist matrix expressed in terms of $\delta$'s is
\be
\hspace{-0.1cm}
\textbf{A}=\frac12\matr{c|ccccc}
{
 0&\de_3-\de_1&2\de_1-2\de_3&\de_3-\de_1&2\de_2&-2\de_2 \\
\hline
 \de_1-\de_3&0&\de_1-\de_3&\de_3-\de_1&-\de_1-\de_3&\de_1+\de_3 \\
 2\de_3-2\de_1&\de_3-\de_1&0&\de_1-\de_3&2\de_1+2\de_2+2\de_3&-2\de_1-2\de_2-2\de_3 \\
 \de_1-\de_3&\de_1-\de_3 &\de_3-\de_1&0&-\de_1-2\de_2-\de_3&\de_1+2\de_2+\de_3 \\
 -2\de_2&\de_1+\de_3&-2\de_1-2\de_2-2\de_3&\de_1+2\de_2+\de_3 &0&0 \\
 2\de_2&-\de_1-\de_3&2\de_1+2 \de_2+2\de_3&-\de_1-2\de_2-\de_3&0&0 \\
}.
\ee
\subsection{Duality properties}
In this section we investigate the duality properties of the twist charges appearing in the Bethe ansatz equations, which were derived in detail in \cite{Bai:2016pxs}. In that paper a series of relations between twist charges was obtained for orbifold ABJM theories  due to the dynamical and fermionic duality. For $\ga$-deformation of ABJM theory, the case is similar, at least at the level of the Bethe ansatz equations. It turns out that all the relations are nicely satisfied. \\
Let's consider them carefully. The first important relation in $\ga$-deformed ABJM theory takes the form
\be\label{DynaDual1}
(\textbf{AK})_3=(\textbf{AK})_1-(\textbf{AK})_0
\ee
for the $sl(2)$ grading, and
\be\label{DynaDual2}
(\ti{\textbf{A}}\ti{\textbf{K}})_3=(\ti{\textbf{A}}\ti{\textbf{K}})_1+(\ti{\textbf{A}}\ti{\textbf{K}})_0
\ee
for the $su(2)$ grading. Here and in the following $\textbf{K}=(L|K_1 ,K_2 ,K_3 ,K_4 ,K_{\bar4})$;$\textbf{K}$ and $\textbf{K}'$ are defined in analogous way for the $su(2)$ and distinguished gradings, respectively. Here we used a tilde to denote the variables in the $su(2)$ grading. See Appendix~\ref{appenb} for its explicit expression.
Relations (\ref{DynaDual1}),(\ref{DynaDual2}) are due to the dynamic duality property, which is essential for all loop Bethe ansatz equations. They are satisfied in the two gradings separately. They are actually  relations linking elements of the twist matrix since the excitation numbers in both sides of these equations are the same.

The remaining relations are
\be\label{ChargeRelation2}
(\ti{\textbf{A}}\ti{\textbf{K}})_4+(\ti{\textbf{A}}\ti{\textbf{K}})_3=(\textbf{AK})_4,\, (\ti{\textbf{A}}\ti{\textbf{K}})_{\bar 4}+(\ti{\textbf{A}}\ti{\textbf{K}})_3=(\textbf{AK})_{\bar 4}.
\ee
\be\label{ChargeRelation3}
(\ti{\textbf{A}}\ti{\textbf{K}})_0-2(\ti{\textbf{A}}\ti{\textbf{K}})_3
=(\ti{\textbf{A}}\ti{\textbf{K}})_2-(\textbf{AK})_2.
\ee
\be\label{ChargeRelation35}
(\textbf{AK})_0-2(\ti{\textbf{A}}\ti{\textbf{K}})_3=(\ti{\textbf{A}}\ti{\textbf{K}})_2-(\textbf{AK})_2.
\ee
\be\label{ChargeRelation4}
(\ti{\textbf{A}}\ti{\textbf{K}})_1+(\textbf{AK})_1=0.
\ee
\be\label{ChargeRelation5}
(\ti{\textbf{A}}\ti{\textbf{K}})_3+(\textbf{AK})_3=0.
\ee

They are essential for the equivalence of the two different gradings and are all satisfied if we take the change of both twist matrix and excitation numbers into account when we switch from one grading to the other.
Recall that the excitation numbers are related as \footnote{Notice that in the undeformed ABJM theory, generically the non-trivial relations among the excitation numbers are $\ti{K}_1=K_2-K_1-1, \ti{K}_3=K_2+K_4+K_{\bar 4}-K_3-1$. As we stated in the main text,  for generically deformed theory with non-trivial phases for the $1$st and the $3$rd Bethe equations, the two $-1$'s will disappear. This can be seen by comparing the two proofs  for fermionic duality in \cite{c9} and \cite{Bai:2016pxs}.}
\bea
&&\ti{K}_1=K_2-K_1,\nn\\
&&\ti{K}_3=K_2+K_4+K_{\bar4}-K_3,\nn\\
&&\ti{K}_i=K_i,  \quad i=2,4,\bar4.
\eea
\par The distinguished grading can not be used for all loop ABAs, so we will not discuss dynamical duality in this grading.
 However this grading can be used at two loop order.
This grading is related to the $su(2)$ grading by fermionic duality. The relations among excitation numbers are
\bea \tilde{K}_1&=&K^\prime_3-K^\prime_2,\nn \\
\tilde{K}_2&=&K^\prime_1+K^\prime_3-K^\prime_2,\nn \\
\tilde{K}_3&=&K^\prime_3,\nn \\
\tilde{K}_4&=&K^\prime_4,\nn \\
\tilde{K}_{\bar 4}&=&K^\prime_{\bar 4}.
 \eea
 Here we used a prime to denote the variables in the distinguished grading.
 From the fermionic duality, we obtain the following relations
 \bea (\tilde{\textbf{A}}\tilde{\textbf{K}})_1&=&-(\textbf{A}^\prime \textbf{K}^\prime)_1-(\textbf{A}^\prime \textbf{K}^\prime)_2,\nn\\
(\tilde{\textbf{A}}\tilde{\textbf{K}})_2&=&(\textbf{A}^\prime \textbf{K}^\prime)_1,\nn\\
(\tilde{\textbf{A}}\tilde{\textbf{K}})_3&=&(\textbf{A}^\prime \textbf{K}^\prime)_2+(\textbf{A}^\prime \textbf{K}^\prime)_3,\nn\\
(\tilde{\textbf{A}}\tilde{\textbf{K}})_4&=&(\textbf{A}^\prime \textbf{K}^\prime)_4,\nn\\
(\tilde{\textbf{A}}\tilde{\textbf{K}})_{\bar 4}&=&(\textbf{A}^\prime \textbf{K}^\prime)_{\bar4}.
 \eea
It is not hard to confirm that all these relations are valid.
\section{Asymptotic Bethe ansatz}\label{s2}
Having obtained the twist matrix, it is straightforward to write down the asymptotic Bethe ansatz equation. We will mainly work in the $sl(2)$ favored grading. Let's first introduce some notation useful in the following.
First we define the Zhukovski variable $x$ through
\be
x+\frac1x=\frac{u}{h(\lambda)},
\ee
where $h(\lambda)$ is the so-called interpolating function \cite{GGY}-\cite{NT} which plays the role of effective coupling in the Bethe ansatz and TBA. Unlike the $\ml{N}=4$ SYM, it has nontrivial dependence on $\lambda$. It has the following behavior at weak coupling \cite{Minahan:2009aq}-\cite{Leoni:2010tb},
\be
h(\lambda)=\lambda-\frac{\pi^2}{3}\lambda^3+\ml{O}(\lambda^5),
\ee
and strong coupling \cite{McLoughlin:2008he}-\cite{Bianchi:2014ada},
\be
h(\lambda)=\sqrt\frac{\lambda}{2}-\frac{\log 2}{2\pi}-\frac{1}{48\sqrt{2\lambda}}+\ml{O}((\frac{1}{\sqrt{\lambda}})^2).
\ee
Recently its exact form has been conjectured in \cite{Gromov} by comparing the quantum spectral
curve method \cite{Cavaglia:2014exa} and supersymmetric localization. The ``physical'' and ``mirror'' branch of the function $x(u)$ are defined as
\be\label{phyandmir}
x^{\text{ph}}(u)=\frac12\(\frac{u}{h}+\sqrt{\frac{u}{h}-2}\sqrt{\frac{u}{h}+2}\),
x^{\text{mir}}(u)=\frac12\(\frac{u}{h}+i\sqrt{4-\frac{u^2}{h^2}}\)\,.
\ee
In this section we will use $x(u)$ to mean $x^{\text{ph}}(u)$. The energy and momentum for a single Bethe root $u_4$ and $u_{\bar4}$ are given by
\be
\ep=\frac12+h(\lambda)\(\frac{i}{x^+}-\frac{i}{x^-}\), p=\frac{1}{i}\log\frac{x^+}{x^-},
\ee
and the total momentum corresponding to the first conserved charge $\ml{Q}_1$ is
\be
\ml{Q}_1=\sum_{j=1}^{K_4}p(u_{4,j})+\sum_{j=1}^{K_{\bar4}}p(u_{\bar4,j}).
\ee
Using the notation of \cite{Gromov:2010dy}, the Bethe equations have the form
\bea\label{ABA}
&&e^{-2\pi i(\textbf{AK})_1}=e^{i\ml{Q}_1}\frac{Q_2^-B^{(+)}}{Q_2^+B^{(-)}}\bigg|_{u_{1,k}}\;,\nn\\
&&e^{-2\pi i(\textbf{AK})_2}=-\frac{Q_2^{++}Q_1^{-}Q_{3}^-}{Q_2^{--}Q_1^{+}Q_{3}^+}\bigg|_{u_{2,k}}\;,\nn\\
&&e^{-2\pi i(\textbf{AK})_3}=\frac{Q_2^-R^{(+)}}{Q_2^+R^{(-)}}\bigg|_{u_{3,k}}\;,\nn\\
&&e^{-2\pi i(\textbf{AK})_4}\(\frac{x_{4,k}^+}{x_{4,k}^-}\)^L=
\frac{B_1^{+}R_3^+B_4^{(+)+}R_{\bar4}^{(-)-}}{B_1^{-}R_3^-B_4^{(-)-}R_{\bar4}^{(+)+}}
\(\prod_{j=1}^{K_4}\frac{x_{4,j}^+}{x_{4,j}^-}\)S\bigg|_{u_{4,k}}\;,\nn\\
&&e^{-2\pi i(\textbf{AK})_{\bar4}}\(\frac{x_{\bar4,k}^+}{x_{\bar4,k}^-}\)^L=
\frac{B_1^{+}R_3^+B_{\bar4}^{(+)+}R_{4}^{(-)-}}{B_1^{-}R_3^-B_{\bar4}^{(-)-}R_{4}^{(+)+}}
\(\prod_{j=1}^{K_{\bar4}}\frac{x_{\bar4,j}^+}{x_{\bar4,j}^-}\)S\bigg|_{u_{\bar4,k}}\;,
\eea
where various functions above are defined as\footnote{Here we use the same definitions as \cite{Gromov:2010dy}, which are slightly different from \cite{Gromov:2009at}. This explains the right hand side of our ABA eqs.~(\ref{ABA})are slightly different from \cite{Gromov:2009at}.}:
\bea
&&R_l^{(\pm)}=\prod_{j=1}^{K_l}\(x(u)-x^{\mp}_{l,j}\)\,,R_l=\prod_{j=1}^{K_l}\(x(u)-x_{l,j}\)\,,\nn\\
&&B_l^{(\pm)}=\prod_{j=1}^{K_l}\(\frac{1}{x(u)}-x^{\mp}_{l,j}\)\,,B_l=
\prod_{j=1}^{K_l}\(\frac{1}{x(u)}-x_{l,j}\)\,,\nn\\
&&Q_l=\prod_{j=1}^{K_l}(u-u_{l,j})\,,S_l=\prod_{j=1}^{K_l}\sigma_{BES}\(x(u),x_{l,j}\)\,,
\eea
and the functions with no index mean a product of type-$4$ and type-$\bar4$ ones: $R=R_4R_{\bar4},B=B_4B_{\bar4},S=S_4S_{\bar4}$.
We have used the general notation
\be
\quad f^{[\pm a]}\equiv f(u\pm ia/2),f^{\pm}\equiv f(u\pm i/2),f^{\pm\pm}\equiv f(u\pm i).
\ee
The Bethe roots must additionally be constrained by the momentum condition
\be
\prod_{j=1}^{K_4}\frac{x^+_{4,j}}{x^-_{4,j}}\prod_{j=1}^{K_{\bar4}}
\frac{x^+_{\bar4,j}}{x^-_{\bar4,j}}=e^{-2\pi i(\textbf{AK})_0}\Leftrightarrow \ml{Q}_1=2\pi m-2 \pi(\textbf{AK})_0,
\ee where $m$ is an integer. The anomalous dimension of the single-trace operator is given by
\be E=h(\lambda)(\sum_{j=1}^{K_4}(\frac{i}{x_{4, j}^+}-\frac{i}{x_{4, j}^-})+\sum_{j=1}^{K_{\bar{4}}}(\frac{i}{x_{\bar{4}, j}^+}-\frac{i}{x_{\bar{4}, j}^-})).
\ee

\section{ The $\ga-$deformed $\text{AdS}_4/\text{CFT}_3$ Y-system}\label{s3}
\begin{figure}
\centering
\includegraphics[width=8cm]{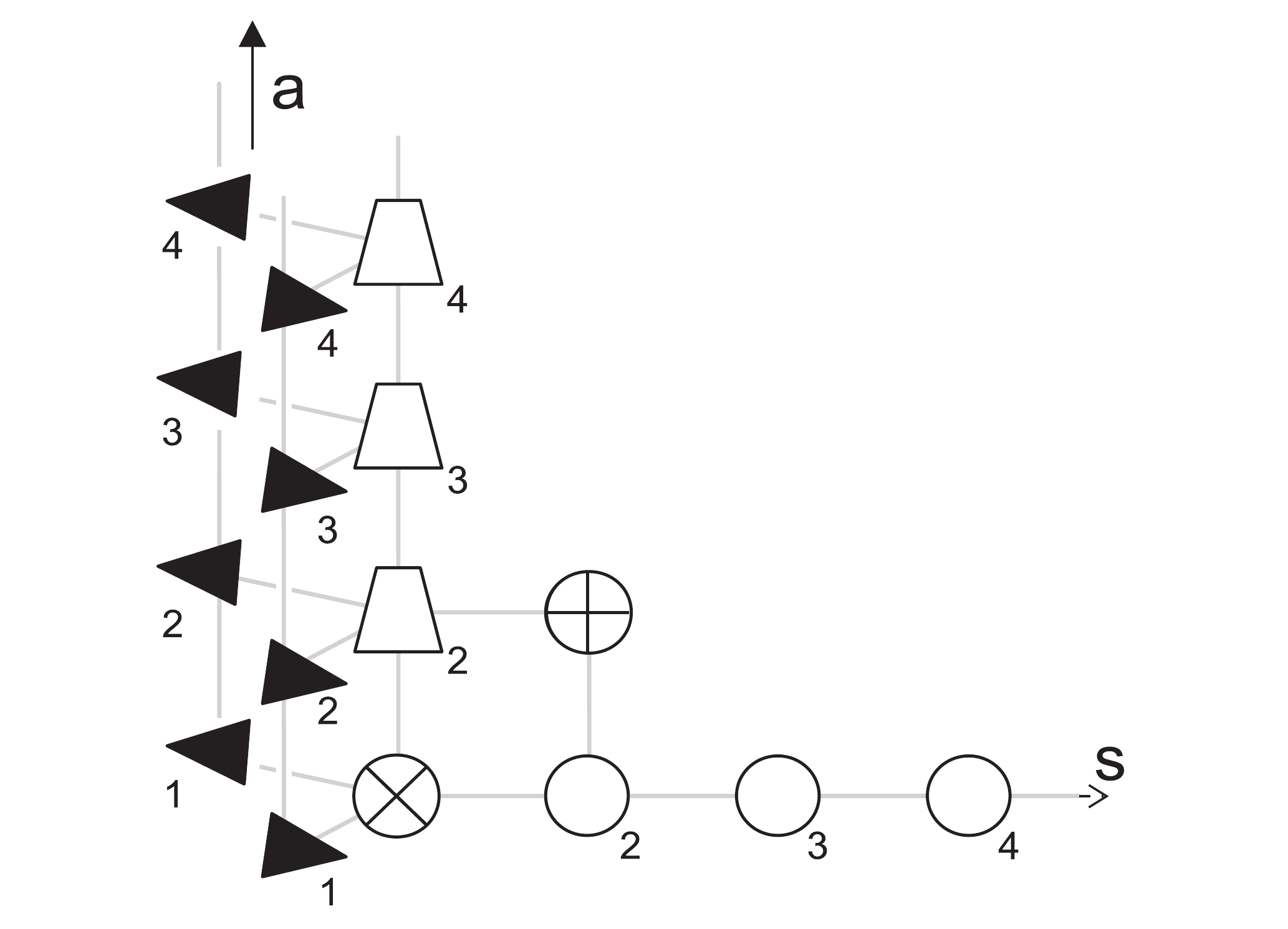}
\caption{The $\text{AdS}_4/\text{CFT}_3$ Y-system. As shown in the figure, we have considered the most generic case when the two black node Y-functions are not equal to each other. We denote them as $Y_{\blacktriangleright_a}$ and $Y_{\blacktriangleleft_a}$ in the main body.}\label{fig}
\end{figure}

In this section, we briefly review some basic facts about the $\text{AdS}_4/\text{CFT}_3$ Y-system, for a thorough treatment see \cite{Gromov:2009at,Bombardelli:2009xz}. The exact spectral information of the AdS/CFT correspondence could be computed through Y-functions which are constrained by the Y-system equations. Because we twist our theory, the TBA equations are slightly changed by adding some appropriate chemical potentials. However the Y-system is the same as the one for the untwisted theory. The equations of a generic Y-system basically have the form
\be
Y_M^+Y_M^-=\frac{\prod_H(1+Y_H)}{\prod_V(1+1/Y_V)},
\ee
where $Y_M^{\pm}=Y_M(u\pm i/2)$ and the index $H$($V$) represent the nodes nearest to the $M$ node in the horizontal(vertical) direction in the  Fig.~\ref{fig}.
The explicit expressions of $\text{AdS}_4/\text{CFT}_3$ Y-system equations have some unusual form, however they are not relevant here. Once the Y-functions are found, the exact energy of a state can be expressed by
\be\label{energy}
E=\sum_{j=1}^{K_4}\ep_1^{\text{ph}}(u_{4,j})+\sum_{j=1}^{K_{\bar4}}\ep_1^{\text{ph}}(u_{\bar4,j})+\de E\;,
\de E=\sum^{\infty}_{a=1}\int_{-\infty}^{\infty}\frac{du}{2\pi i}\frac{\partial\ep_a^{\text{mir}}(u)}{\partial u}
\log(1+Y^{\text{mir}}_{\blacktriangleright_a})(1+Y^{\text{mir}}_{\blacktriangleleft_a}),
\ee
where the rapidities $u_{4,j}$ and $u_{\bar4,j}$ are fixed by the exact Bethe ansatz equations
\be\label{ExactBE}
Y^{\text{ph}}_{\blacktriangleleft_1}(u_{4,j})=-1\;,\quad Y^{\text{ph}}_{\blacktriangleright_1}(u_{\bar4,j})=-1\;.
\ee
$\ep_{n}$ is the asymptotic energy of a physical $n$-magnon bound state when evaluated in the physical kinematics and defines the asymptotic momenta of mirror bound states when evaluated in the mirror kinematics:
\be\label{magnonenergy}
\ep_n(u)=\frac{n}{2}+h(\lambda)\(\frac{i}{x^{[+n]}}-\frac{i}{x^{[-n]}}\).
\ee
The index `$\text{ph}$' and `$\text{mir}$' labeled on functions $\ep_a(u)$, $Y_{\blacktriangleright_a}(u)$ and $Y_{\blacktriangleleft_a}(u)$ in eq.~(\ref{energy}) and eq.~(\ref{ExactBE}) are evaluated at physical and mirror kinematics defined in eq.~(\ref{phyandmir}) respectively. Eq.~(\ref{magnonenergy}) is valid for both physical and mirror kinematics.

\subsection{The twisted generating functional}
Our asymptotic Bethe ansatz equations can be derived from the twisted $\text{AdS}_4/\text{CFT}_3$ Y-system. Inspired by \cite{Gromov:2010dy}, we propose a twisted generating functional $\ml{W}$ which generates an infinite set of transfer matrix eigenvalues of symmetric and anti-symmetric irreducible representations
$T_{1, s}, T_{a, 1}$  as
\be
\ml{W}=\sum_{s=0}^{\infty}T_{1,s}^{[1-s]}D^s\;,\qquad\ml{W}^{-1}
=\sum_{a=1}^{\infty}(-1)^aT_{a,1}^{[1-a]}D^a\,,
\ee
where $D=e^{-i\partial_u}$.
The generating functional we propose in the $sl(2)$ favored grading is
\be
\ml{W}=\(1-\tau_1\frac{B^{(+)+}Q_1^{-}R^{(+)-}}{B^{(-)+}Q_1^{+}R^{(-)-}}D\)
\frac{1}{\(1-\tau_2\frac{Q_1^{-}Q_2^{++}R^{(+)-}}{Q_1^+Q_2R^{(-)-}}D\)}
\frac{1}{\(1-\frac{1}{\tau_2}\frac{Q_2^{--}Q_3^{+}R^{(+)-}}{Q_2Q_3^-R^{(-)-}}D\)}
\(1-\frac{1}{\tau_1}\frac{Q_3^{+}}{Q_3^{-}}D\).
\ee
Then we find
\be\label{Transfer}
T_{1,1}=\frac{R^{(+)-}}{R^{(-)-}}\[\tau_2\frac{Q_1^{-}Q_2^{++}}{Q_1^+Q_2}
-\tau_1\frac{B^{(+)+}Q_1^{-}}{B^{(-)+}Q_1^{+}}
+\frac{1}{\tau_2}\frac{Q_2^{--}Q_3^{+}}{Q_2Q_3^-}
-\frac{1}{\tau_1}\frac{Q_3^{+}R^{(-)-}}{Q_3^{-}R^{(+)-}}\].
\ee
The analyticity of $T_{1,1}$ at zeros of the denominators will give the asymptotic Bethe equations for $u_1,u_2,u_3$
\be\label{u123}
\frac{\tau_1}{\tau_2}\frac{B^{(+)}Q_2^-}{B^{(-)}Q_2^+}\bigg|_{u_{1,k}}=1\;,
\tau_2^2\frac{Q_2^{++}Q_1^{-}Q_{3}^-}{Q_2^{--}Q_1^{+}Q_{3}^+}\bigg|_{u_{2,k}}=-1\;,
\frac{\tau_1}{\tau_2}\frac{Q_2^-R^{(+)}}{Q_2^+R^{(-)}}\bigg|_{u_{3,k}}=1\;.
\ee
We see that if we identify the twists $\tau_1,\tau_2$ as
\be
\begin{aligned}
&\tau_1=e^{i\frac{\pi}{2}((\de_1-\de_3)(K_1-2K_2+K_3)-\de_2(2K_4-2K_{\bar4}))},\\
&\tau_2=e^{i\frac{\pi}{2}((\de_3-\de_1)(2L+K_1-K_3)+(\de_1+\de_2+\de_3)(2K_4-2K_{\bar4}))},
\end{aligned}
\ee
we find exactly our Bethe equations for $u_1,u_2,u_3$ in the $sl(2)$ grading.
Notice that the consistence of the first and the third
equations in (\ref{u123}) is guaranteed by eq.~(\ref{DynaDual1}), which also plays an important role for  the validity of the dynamical duality transformation.
\subsection{Consistency checks}
In this section we perform some consistency checks along the lines of  \cite{Arutyunov:2010gu, deLeeuw:2012hp}. We calculate the twists appearing in the total momentum and the twisted generating functional through the twisted boundary condition of fields eq.~(\ref{TwistedBC}). In order to do this, we had better to rewrite the twist matrix $\textbf{A}$ in terms of the three conserved angular momentum instead of the excitation numbers. In the $sl(2)$ grading, we have
\bea
&&J_1=\frac{p_1}{2}=\frac12(L-K_4+K_{\bar4}+K_2-K_3),\nn\\
&&J_2=\frac{p_2}{2}=\frac12(L+K_4-K_{\bar4}+K_2-K_3),\nn\\
&&J_3=q+\frac{p_1+p_2}{2}=L+K_1-K_2.
\eea
Notice that for the $sl(2)$ grading, the relation between $[p_1, q, p_2]$ and the Dynkin labels $r_j, j=1, \cdots, 4, \bar{4}$ of $OSp(6|4)$ is(see appendix B in \cite{Chen:2016niy} for more information)
\bea
p_1&=&r_3+r_4,\nn\\
q&=&r_2,\nn\\
p_2&=&r_3+r_{\bar 4}.
\eea
The total momentum of the spin chain reads
\be
P=2\pi m-2\pi(\textbf{AK})_0=2\pi m-\pi(\ga_1(2J_2-J_3)+\ga_2(J_3-2J_1)+\ga_3(J_1-J_2)).
\ee
Since we choose the vacuum as $\text{Tr}(Y^1Y_4^{\dg})^L$, the twisted boundary condition for the combined fields $Y^1Y_4^{\dg}$  gives the total momentum
\be
Y^1Y_4^{\dg}(2\pi)=Y^1Y_4^{\dg}(0)e^{i(\de\phi_3+\frac{\de\phi_1+\de\phi_2}{2})}=e^{iP}
\ee
as we expected. Here and in the following, $\de\phi_i=-2\pi\ep_{ijk}\ga_jJ_k$.
Let us make some comments on these phases $\tau_1,\tau_2$, whose origin has a very nice explanation. In literature \cite{Arutyunov:2010gu, deLeeuw:2012hp}, the transfer matrix seems to have only one non-trivial phase for $\ga$-deformed $\text{AdS}_5/\text{CFT}_4$. But we have two here. What is the problem? In fact, if we notice that $\tau_1$ can be expressed using $x^\pm_{4, j}$ and $x^\pm_{\bar4, j}$
\be
\tau_1=e^{-\pi i(\textbf{AK})_0}=\prod_{j=1}^{K_4}\sqrt{\frac{x^+_{4,j}}{x^-_{4,j}}}
\prod_{j=1}^{K_{\bar4}}\sqrt{\frac{x^+_{\bar4,j}}{x^-_{\bar4,j}}},
\ee
there is only one apparent phase $\tau_2$ left to be explained. The global symmetry $OSp(6|4)$ of $\text{AdS}_4/\text{CFT}_3$ breaks down to centrally extended $SU(2|2)$ after we choose the vacuum. We can insert some integrable twist $G\in SU(2)_r\times SU(2)_G\subset SU(2|2)$ into the transfer matrix \cite{deLeeuw:2012hp}. Consider a group element of the form
\be
G=\text{diag}(e^{i\varphi_1},e^{-i\varphi_1},e^{i\varphi_2},e^{-i\varphi_2})\in SU(2)_r\times SU(2)_G
\ee
For the $\ga$-deformation,  we set $\varphi_1=0$ since there are no deformations in  the $AdS_4$ part of the string theory background. Then only $e^{i\varphi_2}$ is relevant and this is consistent with that  $\text{diag}(e^{i\varphi_2},e^{-i\varphi_2})\in SU(2)_G$ is the unbroken part of $SU(4)_R$.
The $SU(2)_G$ transforms $Y^2\rightarrow Y^3$ or $Y_3^{\dg}\rightarrow Y_2^{\dg}$. This suggests us to consider the twisted boundary condition of the combined fields
\be
Y_2^{\dg}Y^3(2\pi)=Y_2^{\dg}Y^3(0)e^{i(\frac{\de\phi_1+\de\phi_2}{2}-\de\phi_3)}\equiv Y_2^{\dg}Y^3(0)e^{i\psi}
\ee
After rewriting $\tau_2$ in terms of the three conserved angular momentum $J_1,J_2,J_3$ as
\be
\tau_2=e^{i(\frac{\pi}{2}(-\ga_1(2J_2+J_3)+\ga_2(2J_1+J_3)+\ga_3(J_1-J_2)))},
\ee
we find the relation
\be
\tau_2=e^{-i\varphi_2}=e^{-i\frac{\psi}{2}},
\ee
which confirms our proposal.
\subsection{The asymptotic solution}
The Bethe equations for momentum-carrying $u_4,u_{\bar4}$ can be found from the exact Bethe equations (\ref{ExactBE}).
The asymptotic large $L$ solutions of the black nodes are given by \cite{Gromov:2009at}
\bea
Y_{\blacktriangleleft_a}\simeq\(\frac{x^{[-a]}}{x^{[+a]}}\)^LT_{a,1}\prod_{n=-\frac{a-1}{2}}^{\frac{a-1}{2}}
\Phi^{\theta_{na}^{E}}_4(u+in)\Phi^{\theta_{na}^{O}}_{\bar4}(u+in)\;,\\
Y_{\blacktriangleright_a}\simeq\(\frac{x^{[-a]}}{x^{[+a]}}\)^LT_{a,1}\prod_{n=-\frac{a-1}{2}}^{\frac{a-1}{2}}
\Phi^{\theta_{na}^{O}}_4(u+in)\Phi^{\theta_{na}^{E}}_{\bar4}(u+in)\;,
\eea
where $\theta_{na}^{E}$ is 0 for even and 1 for odd factors in the product,
\be
\theta_{na}^E\equiv\left\{\begin{array}{ll}
1,&\text{$n+\frac{a-1}{2}$ is even}\\
0,&\text{$n+\frac{a-1}{2}$ is odd}
\end{array}\right.
\ee
and $\theta_{na}^O\equiv1-\theta_{na}^E$.
To match the ABA equations for $u_4$ and $u_{\bar4}$, we have to deform the asymptotic solutions of the two black node Y-functions by the phases $\tau_0$ and $\tau_{\bar0}$, respectively.
\be
\Phi_4(u)=\frac{B_4^{(+)+}R_{\bar4}^{(-)-}B_1^+B_3^-}{B_4^{(-)-}R_{\bar4}^{(+)+}B_1^-B_3^+}
\(\prod_{j=1}^{K_4}\frac{x_{4,j}^+}{x_{4,j}^-}\)S\tau_0\;,\quad
\Phi_{\bar4}(u)=\frac{B_{\bar4}^{(+)+}R_{4}^{(-)-}B_1^+B_3^-}{B_{\bar4}^{(-)-}R_{4}^{(+)+}B_1^-B_3^+}
\(\prod_{j=1}^{K_{\bar4}}\frac{x_{\bar4,j}^+}{x_{\bar4,j}^-}\)S\tau_{\bar0},
\ee
where
\be
\begin{aligned}
\tau_0=e^{i(\frac{3\pi}2\de_1(K_1-2K_2+K_3)-\pi\de_2(2L+2K_2-2K_3+K_4-K_{\bar4})
+\frac{\pi}2\de_3(K_1-2K_2+K_3))},\\
\tau_{\bar0}=e^{i(-\frac{\pi}2\de_1(K_1-2K_2+K_3)+\pi\de_2(2L+2K_2-2K_3-K_4+K_{\bar4})
-\frac{3\pi}2\de_3(K_1-2K_2+K_3))}.
\end{aligned}
\ee

\section{Applications to $sl(2)$-like sector}\label{s4}
In this section we will apply our proposals and results discussed in the previous sections to the $sl(2)$-like sector. The Y-functions simplify a lot in this sector, where we only excite symmetrically the same number of moment carrying roots $u_4$ and $u_{\bar4}$: $u_{4,j}=u_{\bar4,j}$. This leads to $Y_{\blacktriangleleft_a}= Y_{\blacktriangleright_a}\equiv Y_{a,0}$, so $\tau_0=\tau_{\bar0}$ and then $\de_2=0$. We then have $(\textbf{AK})_0=0, \tau_1=1$ and generically $\de_1\neq\de_3$. From this, we know that  the  two loop Bethe ansatz equations for the twist operators  is the same as the ones in the undeformed theory. The solutions of these equations have been studied in  \cite{Zwiebel:2009vb, Beccaria:2009ny, Beccaria:2010kd}. The deformation affects the results only through $\tau_2$ which appears in $T_{a,1}$.
The exact energy can be expressed as
\be
E=2\sum_{j=1}^{K_4}\ep_1^{\text{ph}}(u_{4,j})+\de E,\,\de E=2\sum_{a=1}^{\infty}\int_{-\infty}^{\infty}\frac{du}{2\pi i}\frac{\partial\ep_a^{\text{mir}}}{\partial u}\log(1+Y_{a,0}^{\text{mir}})\,.
\ee
In this case, the only non-trivial Baxter polynomial is $Q_4=Q_{\bar4}=\prod_{j=1}^{K_4}(u-u_{4,j})$. Then the twisted generating functional $\ml{W}^{-1}$ simplifies as
\be
\begin{aligned}
&\sum_{a=0}^{\infty}(-1)^aT_{a,1}^{[1-a]}D^a=\\
&(1-D)^{-1}\(1-\tau_2\(\frac{R_4^{(+)-}}{R_4^{(-)-}}\)^2D\)
\(1-\frac{1}{\tau_2}\(\frac{R_4^{(+)-}}{R_4^{(-)-}}\)^2D\)
\(1-\(\frac{B_4^{(+)+}R_4^{(+)-}}{B_4^{(-)+}R_4^{(-)-}}\)^2D\)^{-1}.
\end{aligned}
\ee
The asymptotics of the $Y_{a,0}$ functions are then given by
\be
Y_{a,0}(u)\simeq \(\frac{x^{[-a]}}{x^{[+a]}}\)^L\Phi_a(u)T_{a,1}(u),
\ee
where the scalar factor $\Phi_a$ is
\be
\Phi_a(u)=\prod_{k=-\frac{a-1}{2}}^{\frac{a-1}{2}}\Phi(u+ik)\,,\;
\Phi(u)=\frac{B_4^{(+)+}R_4^{(-)-}}{B_4^{(-)-}R_4^{(+)+}}S_4^2\(\prod_{j=1}^{K_4}\frac{x_{4,j}^+}{x_{4,j}^-}\).
\ee
Finally, the leading wrapping corrections at weak coupling is
\be\label{LeadingWrapping}
\de E^{\text{LO}}\simeq-\sum_{a=1}^{\infty}\int\frac{du}{\pi}Y^{\text{mir}}_{a,0}(u)\;.
\ee
To compute the leading wrapping effect, in eq.~(\ref{LeadingWrapping}) all we need is to find the large-volume asymptotics of $Y_{a,0}$ functions evaluated in mirror dynamics.
Here we give some useful formulas:\\
\underline{\emph{Dispersion}}\\
The universal factor $(x^{[-a]}/x^{[+a]})^L$, which is also called kinematic factor in literature, evaluated in the mirror dynamics is
\be
\[\frac{4h^2(\lambda)}{a^2+4u^2}\]^L.
\ee
\underline{\emph{Twisted $T_{a,s}$}}\\
After some computations, one find the following compact formula for $T_{a,1}$\footnote{This result and the one just below are valid when the Bethe roots are symmetric with respect to the origin. This condition is valid for all cases considered in this section.
}
\be
T_{a,1}^{\text{mir}}\simeq 4\sin^2\frac{\pi(\de_1-\de_3)L}{2}(-1)^a\sum_{\substack{p=-a+1\\\Delta p=2}}^{a-1}\(\frac{Q_4^{[p]}}{Q_4^{[1-a]}}\)^2\,.
\ee
\underline{\emph{Fused scalar factor}}\\
To compute the fused scalar factor, one should be careful on the dressing phase, which is evaluated in the mirror-physical kinematics and has different weak coupling expansion from the physical-physical kinematics \cite{Gromov:2009bc}. Taking this into account, we find\footnote{We thank Fedor Levkovich-Maslyuk for having clarified this point to us.}
\be
\Phi_a^{\text{mir}}\simeq\[Q_4^+(0)\]^2\frac{Q_4^{[1-a]}}{Q_4^{[-1-a]}Q_4^{[a-1]}Q_4^{[a+1]}}.
\ee

The asymptotic Bethe equations for states in the $sl(2)$ sector (with twist $L$ and excitations $N$) read
\be
\(\frac{x_k^+}{x_k^-}\)^L=-\prod_{j\neq k}^N\frac{u_k-u_j+i}{u_k-u_j-i}\(\frac{x_k^--x_j^+}{x_k^+-x_j^-}\)^2\sigma_{BES}^2,
\ee
where we denote $u_{4,j}(=u_{\bar4,j})$ as $u_j$.
Anomalous dimensions of twist operators will be expanded in power of $h(\lambda)$ as
\be
\Delta_{L,N}=L+N+\sum_{l\geq1}\gamma_{L,N}^{(l)}h^l(\lambda).
\ee
\subsection{Two-loop wrapping corrections to twist-1 operators}
The twist-1 operators in ABJM theory have  also been discussed in \cite{Zwiebel:2009vb, Beccaria:2009ny, Beccaria:2010kd}. In the undeformed ABJM  theory, wrapping corrections appear at four-loop. But in our case, the wrapping effect appears at two-loop.
The leading order Bethe ansatz equations are
\be\label{bae}
\frac{u_k+\frac{i}{2}}{u_k-\frac{i}{2}}=-\prod_{j\neq k}^{N}\frac{u_k-u_j-i}{u_k-u_j+i},\qquad k,j=1,\dots,N
\ee
It can be written in the very efficient Baxter function formalism
\be
\(u+\frac{i}{2}\)Q(u+i)-\(u-\frac{i}{2}\)Q(u-i)=i(2N+1)Q(u),
\ee
where $Q(u)$ is the Baxter polynomial
\be
Q(u)=\ml{N}\prod_{k=1}^{N}(u-u_k).
\ee
The solution of the Baxter equation (\ref{bae}) is  \cite{Zwiebel:2009vb, Beccaria:2009ny, Beccaria:2010kd}
\be
Q(u)=\setlength\arraycolsep{1pt}
{}_2 F_1\left(\begin{matrix}-N,& &iu+\frac12\\&1 &
 &\end{matrix};2\right).
\ee
Inserting it in the Y-system equations for wrapping effects, we obtain the simple result
\be
\de E^{\text{LO}}=(-1)^N\frac{4}{2N+1}h^2(\lambda)\sin^2\frac{\pi(\de_1-\de_3)}{2}.
\ee
\subsection{Four-loop wrapping corrections to twist-2 operators}
After some manipulation, it turns out that the two loop Bethe equations for twist-2 operators is equivalent to the following leading order Baxter equation \cite{Zwiebel:2009vb, Beccaria:2009ny, Beccaria:2010kd}
\be
\(u+\frac{i}{2}\)^2Q(u+i)-\(u-\frac{i}{2}\)^2Q(u-i)=i(2N+2)uQ(u),
\ee
and its solution is known for even $N$ \cite{Beccaria:2009ny, Beccaria:2010kd}
\be
Q(u)=\setlength\arraycolsep{1pt}
{}_3 F_2\left(\begin{matrix}-\frac{N}{2},& &iu+\frac12,&-iu+\frac12\\& &1,&
1& &\end{matrix};1\right).
\ee
The leading wrapping correction appears at four loops.
\be\label{Twist2Wrapping}
\de E^{\text{LO}}=4\sin^2(\pi(\de_1-\de_3))\gamma_{2,N}^{(4)}h^4(\lambda).
\ee
The Y-system provides the following result for the first 10 values
\bea
&&\gamma_{2,2}^{(4)}=-\frac{1}{3}+\frac{7\pi^2}{45},\nn\\
&&\gamma_{2,4}^{(4)}=-\frac{11}{36}+\frac{11\pi^2}{105},\nn\\
&&\gamma_{2,6}^{(4)}=-\frac{29}{108}+\frac{3607\pi^2}{45045},\nn\\
&&\gamma_{2,8}^{(4)}=-\frac{1543}{6480}+\frac{30001\pi^2}{459459},\nn\\
&&\gamma_{2,10}^{(4)}=-\frac{647}{3024}+\frac{161\pi^2}{2907},\nn\\
&&\gamma_{2,12}^{(4)}=-\frac{24509}{126000}+\frac{3843467\pi^2}{79676025},\nn\\
&&\gamma_{2,14}^{(4)}=-\frac{4252817}{23814000}+\frac{22835561\pi^2}{533216475},\nn\\
&&\gamma_{2,16}^{(4)}=-\frac{8247539}{49896000}+\frac{251177003\pi^2}{6511704225},\nn\\
&&\gamma_{2,18}^{(4)}=-\frac{225956701}{1466942400}+\frac{5079358441\pi^2}{144559833795},\nn\\
&&\gamma_{2,20}^{(4)}=-\frac{8258864171}{57210753600}+\frac{44175747151\pi^2}{1367758427445}.
\eea
However a closed formula has not been found.\\
\par As it is evident in eq.~(\ref{Twist2Wrapping}), for $\de_1=\de_3$ but not necessarily zero (and
$\de_2=0$ as we are in the $sl(2)$-like sector), the four-loop wrapping corrections due to deformation disappear for twist-2 operators. A similar discussion is also suitable for twist-1 operators.
\section{$\beta$-deformation}\label{s5}
$\beta$-deformation of ABJM theory corresponds to choose the three parameters as $(0,\de_2,0)$, and $2\de_2=-\gamma_3\equiv \beta,\ga_1=\ga_2=0$.  As discussed in \cite{Imeroni:2008cr}, in this case the theory preserves $\ml{N}=2$ supersymmetry.
\par In this case, due to the fact that the phases in the scalar factor do not vanish, the two black node Y-functions are different. The leading wrapping corrections are expressed by
\be\label{BetaLOWrapping}
\de E^{\text{LO}}\simeq-\sum_{a=1}^{\infty}\int_{-\infty}^{\infty}\frac{du}{2\pi}(Y^{\text{mir}}_{\blacktriangleleft_a}
+Y^{\text{mir}}_{\blacktriangleright_a})\;.
\ee
\subsection{One-particle state}\label{s3-5}
In this section, we will discuss the leading wrapping of a single magnon state in $\beta$-deformed theory.
We consider the state with only one type-4 excitation $K_4=1$. The unique momentum carrying root $u_4$ must satisfy the following Bethe equation and cyclicity condition
\be
\(\frac{x_4^+}{x_4^-}\)^L=e^{-\pi iL\beta},\qquad \frac{x_4^+}{x_4^-}=e^{-\pi i\beta}.
\ee
A solution is easily obtained
\be
u_4=-\frac12\cot(\frac{\pi\beta}{2})-2\sin(\pi\beta)h^2(\lambda)+\cdots.
\ee
The twisted generating functional reads
\be
\begin{aligned}
&\sum_{a=0}^{\infty}(-1)^aT_{a,1}^{[1-a]}D^a=\\
&(1-e^{i\pi\beta/2}D)^{-1}\(1-e^{-i\pi\beta/2}\frac{R_4^{(+)-}}{R_4^{(-)-}}D\)
\(1-e^{i\pi\beta/2}\frac{R_4^{(+)-}}{R_4^{(-)-}}D\)
\(1-e^{-i\pi\beta/2}\frac{B_4^{(+)+}R_4^{(+)-}}{B_4^{(-)+}R_4^{(-)-}}D\)^{-1}.
\end{aligned}
\ee
After some manipulations, we find the formula
\be
T^{\text{mir}}_{a,1}\simeq(-1)^{a}\frac{16h^2e^{i\pi(a+1)\beta/2}a(a^2+4u^2-\csc^2(\pi\beta/2) )\sin^3(\pi\beta/2)}{(a^2+4u^2)(i-ia+\cot(\pi\beta/2)+2u)}.
\ee
In this case $\tau_0=e^{-\pi i(2L+1)\beta/2}$,$\tau_{\bar0}=e^{\pi i(2L-1)\beta/2}$, and working carefully with these phases we find
\be
\Phi^{\text{mir}}_{\blacktriangleleft_a}\simeq\left\{\begin{array}{ll}
\frac{e^{-i\pi(a+1)\beta/2}(i-ia+\cot(\pi\beta/2)+2u)}
{\sin(\pi\beta/2)(-ia-i+\cot(\pi\beta/2)+2u)(ia+i+\cot(\pi\beta/2)+2u)}&\text{$a$ is even},\\
\\
\frac{e^{-i\pi\beta(a+2L+1)/2}(i-ia+\cot(\pi\beta/2)+2u)}
{\sin(\pi\beta/2)(-ia-i+\cot(\pi\beta/2)+2u)(ia-i+\cot(\pi\beta/2)+2u)}&\text{$a$ is odd}.
\end{array}\right.
\ee
\be
\hspace{-3cm}
\Phi^{\text{mir}}_{\blacktriangleright_a}\simeq\left\{\begin{array}{ll}
\frac{e^{-i\pi(a+1)\beta/2}}{\sin(\pi\beta/2)(ia-i+\cot(\pi\beta/2)+2u)}&\text{$a$ is even},\\
\\
\frac{e^{-i\pi\beta(a-2L+1)/2}}{\sin(\pi\beta/2)(ia+i+\cot(\pi\beta/2)+2u)}&\text{$a$ is odd}.
\end{array}\right.
\ee
Despite the annoying phases appearing in the T-functions and the fused scalar factors, the sums of the gauge invariant Y-functions are real when the two black nodes are combined.
Their structures are totally different for $a$ even or odd.
When $a$ is an even number
\be\label{EvenY}
\hspace{-14cm}
Y^{\text{mir}}_{\blacktriangleleft_a}+Y^{\text{mir}}_{\blacktriangleright_a}\simeq\nn
\ee
\be
\frac{h^{2L+2}2^{2L+6}a(a^2+4u^2-\csc^2(\pi\beta/2))\sin^4(\pi\beta/2)}{(a^2+4u^2)^{L+1}}
\frac{2+a^2+4u^2-(a^2+4u^2)\cos(\pi\beta)+4u\sin(\pi\beta)}{y_ay_{-a}},
\ee
and for odd $a$
\be\label{OddY}
\hspace{-5cm}
Y^{\text{mir}}_{\blacktriangleleft_a}+Y^{\text{mir}}_{\blacktriangleright_a}
\simeq\frac{-h^{2L+2}2^{2L+7}a(a^2+4u^2-\csc^2(\pi\beta/2))\sin^4(\pi\beta/2)}{(a^2+4u^2)^{L+1}}\times\nn
\ee
\be
\hspace{-0.5cm}
\frac{((a^2+4u^2)\sin^2(\pi\beta/2)+\cos(\pi\beta)+2u\sin(\pi\beta))\cos(\pi L\beta)+(2-2u\cos(\pi\beta)+\sin(\pi\beta))\sin(\pi L\beta)}{y_ay_{-a}},
\ee
where
\be\label{y}
y_a=2+a^2+2a+4u^2-(a^2+2a+4u^2)\cos(\pi\beta)+4u\sin(\pi\beta).
\ee
If we take $\beta=1$, these expressions simplify considerably.
The summation cannot be done directly, because the even term and the odd term have different expressions. However, since the series is absolutely convergent, we can add the even term and odd term separately. Plugging eqs.~(\ref{EvenY}),(\ref{OddY}) and (\ref{y}) into eq.~(\ref{BetaLOWrapping}) and summing the even and odd terms together, we get the following leading wrapping corrections at $h^{2L+2}$.
\bea
&&\de E^{\beta=1}_{L=1}/h^4=-8\zeta_2,\nn\\
&&\de E^{\beta=1}_{L=2}/h^6=-16(2\zeta_2-3\zeta_4),\nn\\
&&\de E^{\beta=1}_{L=3}/h^8=136\zeta_4-155\zeta_6,\nn\\
&&\de E^{\beta=1}_{L=4}/h^{10}=-4(56\zeta_4+75\zeta_6-140\zeta_8).
\eea
In the computations above, we found all the odd powers of $\pi$ are canceled out.
\subsection{A simple case of two particle state}
\par 
In this subsection, we compute the wrapping effect for the operator with $L=K_4=K_{\bar 4}=1$ and other $K$'s being zero.
At lowest order, the $\beta$-deformed Bethe equations read
\bea
\frac{u_4+\frac{i}{2}}{u_4-\frac{i}{2}}=e^{-\pi i\beta}\frac{u_4-u_{\bar4}-i}{u_4-u_{\bar4}+i}\,,\nn\\
\frac{u_{\bar4}+\frac{i}{2}}{u_{\bar4}-\frac{i}{2}}=e^{\pi i\beta}\frac{u_{\bar4}-u_{4}-i}{u_{\bar4}-u_{4}+i}\,,
\eea
and the momentum conservation equation is not deformed
\be
\frac{u_4+\frac{i}{2}}{u_4-\frac{i}{2}}\frac{u_{\bar4}+\frac{i}{2}}{u_{\bar4}-\frac{i}{2}}=1\,.
\ee
Their root is easily found to be $u_4=\Delta,u_{\bar4}=-\Delta$ with $\Delta=\frac12\tan\frac{\pi\beta}{4}$\footnote{There is another root with $\Delta=-\frac12\cot\frac{\pi\beta}{4}$. This solution will not be considered here since it will go to infinity when $\beta\to 0$.}.
In this case, we have two kinds of Baxter polynomial. $Q_{4}(u)=u-\Delta$,$Q_{\bar4}(u)=u+\Delta$ and define $Q=Q_4Q_{\bar4}$. Since $\tau_1=\tau_2=1,\tau_0=e^{-\pi i\beta},\tau_{\bar0}=e^{\pi i\beta}$, the deformation does not appear in the $T$-functions. It only affects the fused scalar factors.
\be
\Phi^{\text{mir}}_{\blacktriangleleft_a}\simeq\left\{\begin{array}{ll}
\frac{Q_4^+(0)Q_{\bar4}^+(0)Q_4^{[1-a]}}{Q_4^{[a+1]}Q_4^{[-a-1]}Q_{\bar4}^{[a-1]}}&\text{$a$ is even},\\
\frac{[Q_{\bar4}^+(0)]^3Q_4^{[1-a]}}{Q^+_{4}(0)Q_4^{[a-1]}Q_4^{[-a-1]}Q_{\bar4}^{[a+1]}}&\text{$a$ is odd}.
\end{array}\right.
\ee
\be
\Phi^{\text{mir}}_{\blacktriangleright_a}\simeq\left\{\begin{array}{ll}
\frac{Q_4^+(0)Q_{\bar4}^+(0)Q_{\bar4}^{[1-a]}}{Q_{\bar4}^{[a+1]}Q_{\bar4}^{[-a-1]}Q_{4}^{[a-1]}}
&\text{$a$ is even},\\
\frac{[Q_{4}^+(0)]^3Q_{\bar4}^{[1-a]}}{Q^+_{\bar4}(0)Q_{\bar4}^{[a-1]}Q_{\bar4}^{[-a-1]}Q_{4}^{[a+1]}}
&\text{$a$ is odd}.
\end{array}\right.
\ee
The $T$-functions are the same as the ones in the undeformed theory
\be
T^{\text{mir}}_{a,1}\simeq i\,c\,h^2(\lambda)\frac{(-1)^{a+1}}{Q^{[1-a]}}\sum_{\substack{p=-a\\\Delta p=2}}^{a}\frac{Q^{[-1-p]}-Q^{[1-p]}}{u-\frac{i}{2}p}\Bigg|_{Q^{[-a-1]},Q^{[a+1]}\rightarrow 0}\;,
\ee
where
\be
c=i(\log Q)'|_{u=-i/2}^{u=i/2}\;.
\ee
For an even $a$
\be
\hspace{-6cm}
Y^{\text{mir}}_{\blacktriangleleft_a}+Y^{\text{mir}}_{\blacktriangleright_a}\simeq
\frac{256h^{4}a \left(a^2-4 \Delta ^2+4 u^2-1\right) }{(a^2+4u^2)^{2}}\times\nn
\ee
\be\label{evenY}
\times\frac{\left(a^4+8 a^2 \Delta ^2+8 a^2 u^2-2 a^2+\left(4 \Delta^2+1\right)^2+16 u^4-32 \Delta ^2 u^2+8 u^2\right)}{\tilde{y}_a\tilde{y}_{-a}},
\ee
and for an odd $a$
\be\label{oddY}
Y^{\text{mir}}_{\blacktriangleleft_a}+Y^{\text{mir}}_{\blacktriangleright_a}\simeq
\frac{-256h^{4}a(a^2-4\Delta^2+4u^2-1)}{(a^2+4u^2)^{2}(1+4\Delta^2)^2}
\frac{\text{num}}{\tilde{y}_a\tilde{y}_{-a}},
\ee
where the numerator is
\bea
&&\text{num}=16a^4\Delta^4-24a^4\Delta^2+a^4+128a^2\Delta^6+32a^2\Delta^4-8a^2\Delta^2\nn\\
&&+128a^2\Delta^4u^2-192a^2\Delta^2u^2+8a^2u^2-2a^2+256\Delta^8+256\Delta^6+96\Delta^4+16\Delta^2\nn\\
&&+256\Delta^4u^4-384 \Delta^2u^4+16 u^4-512\Delta^6u^2-128\Delta^4u^2+32\Delta^2u^2+8u^2+1,
\eea
and
\be
\tilde{y}_a=a^4+4a^3+8a^2\Delta^2+8a^2u^2+6a^2+16a\Delta^2+16au^2+4a+16\Delta^4+8\Delta^2+16u^4-32 \Delta^2u^2+8u^2+1.
\ee
Plugging eq.(\ref{evenY}) and eq.(\ref{oddY}) into the formula for the energy correction eq.~(\ref{BetaLOWrapping}), integrating on $u$ and summing the results, we find
\be
\de E^{\text{LO}}=\frac{8h^4(\lambda)(12-\pi^2(1+4\Delta^2))}{3(1+4\Delta^2)^2}.
\ee
Taking the deformation parameter $\beta\rightarrow0$, which means $\Delta\rightarrow0$,
we get that,
\be
\de E^{\text{LO}}=h^4(\lambda)(32-16\zeta(2)).
\ee
This result is the same as one for the  operator with $L=2, \ti{K}_4=\ti{K}_{\bar 4}=1, \eta=1$ in the undeformed ABJM theory \cite{Gromov:2009tv,Minahan:2009wg} (this operator is in the irreducible  representation $\textbf{20}$ of $SU(4)$). In fact, this operator and the operator we started with are linked by two steps. First we add to the $L=2, \ti{K}_4=\ti{K}_{\bar 4}=1$ operator an $u_3$ root at zero in the $su(2)$ grading and change $L$ from $L=2$ to $L=1$ \cite{Zwiebel:2009vb}. Then we perform the fermionic duality to describe the operator with $L=K_4=K_{\bar 4}=1$ in the $sl(2)$ grading (notice here we are in the undeformed theory, so $K_3=\ti{K}_4+\ti{K}_{\bar 4}-\ti{K}_3-1=0$).

\section{Conclusions}
In this paper we have assumed the integrability of planar $\ga$-deformed ABJM theory at all loop order based on the algebraic Bethe ansatz at two loop in the planar scalar sector, and proposed the asymptotic Bethe ansatz equations via a similar treatment in $\ga$-deformed $\ml{N}=4$ SYM \cite{Beisert:2005if}. Duality properties of the twist charges have been investigated, which are essential for the twisted Bethe equations. Utilising  the Y-system techniques, we compute the leading wrapping corrections for various operators. Some important results of $\beta$-deformed theory are obtained, which should be checked through a direct Feynman diagram computation as in \cite{Gunnesson:2009nn,Fiamberti:2008sm} for $\beta$-deformed SYM. There are strong evidence that planar ABJ theory \cite{ABJ} is integrable as well \cite{Bak:2008vd, Minahan:2009te},\cite{Minahan:2009aq}-\cite{Leoni:2010tb}. As suggested in \cite{Cavaglia:2016ide}, then the results
obtained in this paper on wrapping corrections are also valid for the same quantities in $\gamma$-deformed ABJ theory once having replaced $h(\lambda)$ by the function $h(\lambda_1, \lambda_2)$ given in \cite{Cavaglia:2016ide}. It will be also interesting to find the S-matrix of the spin chain for $\gamma$-deformed ABJM theory based on the studies in \cite{Ahn:2008aa}-\cite{Ahn:2012hs} and to study the integrability of IIA string theory on $AdS_4\times CP^3_\gamma$ on the basis of \cite{Arutyunov:2008if}-\cite{arXiv:1009.3498},\cite{arXiv:0811.1566, arXiv:1101.3777}.

\section*{Acknowledgments}
It is our great pleasure to thank C.~Ahn,  Z.~Chang, F.~Levkovich-Maslyuk, J.~Plefka  and Z.-Y.~Xian for very helpful discussions. We would like to express our special thanks to the anonymous referee for valuable suggestions to improve the paper.   HHC and JW thank Institute of Modern Physics, Northwest University for warm hospitality during the 9th  Summer School in Theoretical Physics on {\it Integrable Models and their Applications}. This work was in part supported by Natural Science Foundation of China under Grant Nos. 11575202(HHC, JW), 11275207(HHC),  11275208(PL), 11575195(PL), 11690022(HHC). JW would also like to thank the participants of the advanced workshop ``Dark Energy and Fundamental Theory'' supported by the Special Fund for Theoretical Physics from NSFC with Grant No.~11447613 for stimulating discussion. JW also gratefully acknowledges the support of K.~C.~Wong Education Foundation.

\begin{appendix}
\section{Algebraic Bethe Ansatz}\label{appena}
The proof of integrability of planar $\ga$-deformed ABJM theory in the scalar sector at two loop level is almost the same as
the one for $\beta$-deformed theory in \cite{He:2013hxd}, the only difference is that now we need to use the star product defined in eq.~(\ref{star}).
So here we only list
the Hamiltonian from the Feynman diagram computations,
\bea  \label{Hamiltonian}\widetilde{H}
&=&\lambda^2\sum_{i=1}^{2L}\left(\mathbb{I}-\widetilde{\mathbb{P}}_{i,
i+2}+\frac12\mathbb{P}_{i, i+2}\mathbb{K}_{i, i+1}
+\frac12\mathbb{P}_{i, i+2}\mathbb{K}_{i+1,
i+2}\right), \eea
where \bea\left(\widetilde{\mathbb{P}}_{i, i+2}\right)^{I_i I_{i+1} I_{i+2}}_{J_i J_{i+1} J_{i+2}}&\equiv&\exp(-2\pi i(Q_{Y^{J_i}}\times Q_{Y^{J_{i+1}}}+Q_{Y^{J_{i+1}}}\times Q_{Y^{J_{i+2}}}+Q_{Y^{J_{i+2}}}\times Q_{Y^{J_i}}))\nonumber\\
&\times&\left(\mathbb{P}_{i, i+2}\right)^{I_i I_{i+1} I_{i+2}}_{J_i
J_{i+1} J_{i+2}},\eea
and the R matrices,
\bea
{\widetilde{{\fR^{\bf 4{4}}}}}(u)^{IJ}_{KL}&=&f^{IJ}_{KL}\(u\mathbb{I}+\mathbb{P}\)^{IJ}_{KL},\label{tr1}\\
\widetilde{{\fR^{\bf 4\bar{4}}}}(u)^{IJ}_{KL}&=&(f^{IJ}_{KL})^{-1}\(-(u+2)\mathbb{I}+\mathbb{K}\)^{IJ}_{KL},\\
\widetilde{{\fR^{\bf \bar{4}4}}}(u)^{IJ}_{KL}&=&(f^{IJ}_{KL})^{-1}\(-(u+2)\mathbb{I}+\mathbb{K}\)^{IJ}_{KL},\\
\widetilde{{\fR^{\bf \bar{4}\bar{4}}}}(u)^{IJ}_{KL}&=&f^{IJ}_{KL}\(u\mathbb{I}+\mathbb{P}\)^{IJ}_{KL},
 \eea
and the definition of $f^{IJ}_{KL}$
is \be f^{IJ}_{KL}=\exp(i\pi(Q_{Y^J}\times Q_{Y^I}-Q_{Y^K}\times Q_{Y^L})). \ee
The obtained Bethe ansatz equations are
\bea \exp\(2\pi i \tilde{Q}\times (Q_{Y^1}-Q_{Y^2})\)\(\frac{u_{4, k}+\frac{i}2}{u_{4, k}-\frac{i}2}\)^L=\mathop{\prod_{j=1}^{K_4}}_{j\ne k}
\frac{u_{4, k}-u_{4, j}+i}{u_{4, k}-u_{4, j}-i}\prod_{j=1}^{K_3}\frac{u_{4, k}-u_{3, j}-\frac{i}{2}}{u_{4, k}-u_{3, j}+\frac{i}{2}}\,, \eea

\bea  \exp\(2\pi i \tilde{Q}\times (Q_{Y^2}-Q_{Y^3})\)=\prod_{j=1}^{K_4}\frac{u_{3, k}-u_{4, j}-\frac{i}2}{u_{3, k}-u_{4, j}+\frac{i}2}
\prod_{j=1}^{K_{\bar{4}}}\frac{u_{3, k}-u_{\bar{4}, j}-\frac{i}2}{u_{3, k}-u_{\bar{4}, j}+\frac{i}2}
\mathop{\prod_{j=1}^{K_3}}_{j\ne k}\frac{u_{3, k}-u_{3, j}+i}{u_{3, k}-u_{4, j}-i}\,,\eea

\bea \exp\(2\pi i \tilde{Q}\times (Q_{Y^3}-Q_{Y^4})\)\(\frac{u_{\bar{4}, k}+\frac{i}2}{u_{\bar{4}, k}-\frac{i}2}\)^L=\mathop{\prod_{j=1}^{K_{\bar{4}}}}_{j\ne k}
\frac{u_{\bar{4}, k}-u_{\bar{4}, j}+i}{u_{\bar{4}, k}-u_{\bar{4}, j}-i}\prod_{j=1}^{K_3}\frac{u_{\bar{4}, k}-u_{3, j}-\frac{i}{2}}{u_{\bar{4}, k}-u_{3, j}+\frac{i}{2}}\,. \eea
Here we have defined \be \tilde{Q}\equiv (L-K_4)Q_{Y^1}+(K_4-K_3)Q_{Y^2}+(K_3-K_{\bar{4}})Q_{Y^3}+(K_{\bar{4}}-L)Q_{Y^4}.\ee
The zero momentum condition is
\be \exp\(2\pi i \tilde{Q}\times (Q_{Y^4}-Q_{Y^1})\)=\prod_{j=1}^{K_4}\frac{u_{4, j}+\frac{i}2}{u_{4, j}-\frac{i}2}
\prod_{j=1}^{K_{\bar{4}}}\frac{u_{\bar{4}, j}+\frac{i}2}{u_{\bar{4}, j}-\frac{i}2}. \ee

These equations can be written as
\be e^{2\pi i(\textbf{AK})_j}\(\frac{u_{j, k}+\frac{i}{2}V_j}{u_{j, k}-\frac{i}{2}V_j}\)^L=\mathop{\prod_{j^\prime=3, 4, \bar{4}}\prod_{k^\prime=1}^{K_{j^\prime}}}_{(j^\prime, k^\prime)\neq (j, k)}\frac{u_{j, k}-u_{j^\prime, k^\prime}+\frac{i}{2}M_{j, j^\prime}}{u_{j, k}-u_{j^\prime, k^\prime}+\frac{i}{2}M_{j, j^\prime}}, \ee
with $j=3, 4, \bar{4}$ and
\be e^{2\pi i(\textbf{AK})_0}=\prod_{j=3, 4, \bar{4}}\prod_{k=1}^{K_j}\frac{u_{j, k}+\frac{i}{2}V_j}{u_{j, k}-\frac{i}{2}V_j}.\ee
Here we used the Cartan matrix of $SO(6)$,
\be M_{jj^\prime}=\matr{ccc}
{2&-1&-1\\
-1&2&0\\
-1&0&2}, \ee
and the Dynkin labels $V_j=(0, 1, 1)$.
Since
we are studying the scalar sector at two-loop level, flipping the signs of $\ga_1, \ga_2, \ga_3$ simultaneously just maps the Hamiltonian to its transpose matrix. Then this  will not change the eigenvalues  of the Hamiltonian.
So the  result here is consistent with the one in the main part of this paper.

In fact, we can  further use the $U(1)_b$ symmetry of ABJM theory as the fourth $U(1)$ symmetry and construct a six-parameter deformation
with the $\mathbf{C}$ matrix being
\bea
\textbf{C}=\matr{cccc}
{
0&-\ga_3&+\ga_2&\alpha_1\\
+\ga_3&0&-\ga_1&\alpha_2\\
-\ga_2&+\ga_1&0&\alpha_3\\
-\alpha_1&-\alpha_2&-\alpha_3&0\\
}.\eea
Now the charges are listed in table~2. Notice that if we choose $\ga_1=\ga_2=\alpha_3=0$, this will go back to the three-parameter deformation considered in \cite{He:2013hxd}.
Using the fact that the four scalars share the same fourth charge, it is not hard to find that $\alpha_1, \alpha_2, \alpha_3$ will not enter the phase of the Hamiltonian (eq.~(\ref{Hamiltonian}))  in the scalar sector at two loop order.

\begin{table}\label{table2}
  \centering
  \begin{tabular}{|c|c|c|c|c|c|c|c|c|}
  \hline
  Fields&$Y^1$&$Y^2$&$Y^3$&$Y^4$&$\psi^{\dg1}$&$\psi^{\dg2}$&$\psi^{\dg3}$&$\psi^{\dg4}$\\
  \hline
  $Q_f^1$&$\frac{1}{2}$&$-\frac{1}{2}$&$0$&$0$ &$\frac{1}{2}$&$-\frac{1}{2}$&$0$&$0$ \\
  \hline
  $Q_f^2$&$0$ & $0$ & $\frac{1}{2}$&$-\frac{1}{2}$&$0$&$0$&$\frac{1}{2}$&$-\frac{1}{2}$\\
  \hline
  $Q_f^3$&$\frac{1}{2}$&$\frac{1}{2}$&$-\frac{1}{2}$&$-\frac{1}{2}$&$\frac{1}{2}$&$\frac{1}{2}$& $-\frac{1}{2}$&$-\frac{1}{2}$ \\
  \hline
  $Q_f^4$&$\frac12$&$\frac12$&$\frac12$&$\frac12$&$-\frac12$&$-\frac12$&$-\frac12$&$-\frac12$\\
  \hline
  \end{tabular}
  \caption{Charges of fields under four $U(1)$ generators.}
\end{table}

\section{The $su(2)$ grading}\label{appenb}
In the appendix, we briefly discuss some relevant formulas in $su(2)$ grading.
The twist matrix is
\be
\textbf{\~{A}}=\frac12\matr{c|ccccc}
 {
 0 &\de_1-\de_3 & 0 & \de_1-\de_3 & -\de_1+2 \de_2+\de_3 & -\de_1-2 \de_2+\de_3 \\
\hline
 \de_3-\de_1 & 0 & 0 & \de_3-\de_1 & 2 \de_1 & -2 \de_3 \\
 0 & 0 & 0 & 0 & 0 & 0 \\
 \de_3-\de_1 & \de_1-\de_3 & 0 & 0 & \de_1+2\de_2+\de_3 & -\de_1-2 \de_2-\de_3 \\
 \de_1-2\de_2-\de_3 & -2\de_1 & 0 & -\de_1-2\de_2-\de_3 & 0 & 2\de_1+4\de_2+2\de_3 \\
 \de_1+2\de_2-\de_3 & 2\de_3 & 0 & \de_1+2\de_2+\de_3 & -2\de_1-4\de_2-2\de_3 & 0 \\
}.
\ee
In this case, the twist matrix still have the form (\ref{Trick}), but with the three different charges
\bea
\textbf{q}_1&=&(0|-1,0,-1,1,1),\nn\\
\textbf{q}_2&=&(1|0,0,1,-2,0),\nn\\
\textbf{q}_3&=&(1|0,0,1,0,-2).
\eea
Here we propose the following twisted generating functional
\be
\ml{W}_{\text{$su(2)$}}=\frac{1}{1-\frac{1}{\tilde{\tau}_1}\frac{B^{(-)-}Q_1^+R^{(-)+}}{B^{(+)-}Q_1^-R^{(-)-}}D}
\(1-\frac{1}{\tilde{\tau}_2}\frac{Q_1^+Q_2^{--}R^{(-)+}}{Q_1^-Q_2R^{(+)+}}D\)
\(1-\tilde{\tau}_2\frac{Q_2^{++}Q_3^-R^{(-)+}}{Q_2Q_3^+R^{(+)+}}D\)\frac{1}{1-\tilde{\tau}_1\frac{Q_3^-}{Q_3^+}D}.
\ee
The cancellation of potential poles in the $T_{a,s}$ functions gives the following Bethe equations
\be
1=\frac{\tilde{\tau}_2}{\tilde{\tau}_1}\frac{Q_2^+B^{(-)}}{Q_2^-B^{(+)}}\bigg|_{u=u_{1,k}}\,,\;
-1=\frac{1}{(\tilde{\tau}_2)^2}\frac{Q_1^+Q_2^{--}Q_3^+}{Q_1^-Q_2^{++}Q_3^-}\bigg|_{u=u_{2,k}}\,,\;
1=\frac{\tilde{\tau}_2}{\tilde{\tau}_1}\frac{Q_2^+R^{(-)}}{Q_2^-R^{(+)}}\bigg|_{u=u_{3,k}}\,.
\ee
We choose
\be
\tilde{\tau}_1=e^{-\pi i((\de_3-\de_1)L+(\de_1-\de_3)\tilde{K}_1+(\de_1+2\de_2+\de_3)(\tilde{K}_4-\tilde{K}_{\bar4}))}
\,,\quad\tilde{\tau}_2=1
\ee
to match the Bethe equations for $u_1,u_2,u_3$ in this grading. The scalar factors should behave as
\be
\tilde{\Phi}_4(u)=\frac{Q_4^{++}B_3^+B_1^-}{Q_4^{--}B_3^-B_1^+}S\tilde{\tau}_0\,,\qquad
\tilde{\Phi}_{\bar4}(u)=\frac{Q_{\bar4}^{++}B_3^+B_1^-}{Q_{\bar4}^{--}B_3^-B_1^+}S\tilde{\tau}_{\bar0}\,,
\ee
where
\be
\ti{\tau}_0=e^{\pi i(-2\de_2L-(\de_1+\de_3)\ti{K}_1-(\de_1+2\de_2+\de_3)(\ti{K}_3-K_4-K_{\bar4}))}\,,
\ti{\tau}_{\bar0}=e^{\pi i(2\de_2L+(\de_1+\de_3)\ti{K}_1+(\de_1+2\de_2+\de_3)(\ti{K}_3-K_4-K_{\bar4}))}\,.
\ee
\end{appendix}

\end{document}